\documentclass[twocolumn,showpacs,preprintnumbers,amsmath,amssymb]{revtex4}

\usepackage{graphicx}
\usepackage{epstopdf}
\usepackage{color}

\begin{document}


\title{ Room temperature dielectric and magnetic properties of Gd and Ti co-doped BiFeO$_{3}$ ceramics}

\author{M. A. Basith}
\affiliation{
Department of Physics, Bangladesh University of Engineering and Technology, Dhaka, Bangladesh.
}
\author{O. Kurni}
\affiliation{
Department of Physics, Bangladesh University of Engineering and Technology, Dhaka, Bangladesh.
}
\author{M. S. Alam}
\affiliation{
Department of Physics, Bangladesh University of Engineering and Technology, Dhaka, Bangladesh.
}
\author{B. L. Sinha}
\affiliation{
Atomic Energy Center, Dhaka, Bangladesh.
}
\author{Bashir Ahmmad}
\affiliation{
Graduate School of Science and Engineering, Yamagata University, 4-3-16 Jonan, Yonezawa 992-8510, Japan.
}


\date{\today}

\begin{abstract}
Room temperature dielectric and magnetic properties of BiFeO$_3$ samples, co-doped with magnetic Gd and non-magnetic Ti in place of Bi and Fe, respectively, were reported. The nominal compositions of Bi$_{0.9}$Gd$_{0.1}$Fe$_{1-x}$Ti$_x$O$_3$ (x = 0.00-0.25) ceramics were synthesized by conventional solid state reaction technique. X-ray diffraction patterns revealed that the substitution of Fe by Ti induces a phase transition from rhombohedral to orthorhombic at x $>$ 0.20. Morphological studies demonstrated that the average grain size was reduced from $\sim {~}$1.5 $\mu m$ to $\sim {~}$200 $nm$ with the increase in Ti content. Due to Ti substitution, the dielectric constant was stable over a wide range of high frequencies (30 kHz to 20 MHz) by suppressing the dispersion at low frequencies.  The dielectric properties of the compounds are associated with their improved morphologies and reduced leakage current densities probably due to the lower concentration of oxygen vacancies in the compositions. Magnetic properties of Bi$_{0.9}$Gd$_{0.1}$Fe$_{1-x}$Ti$_x$O$_3$ (x = 0.00-0.25) ceramics measured at room temperature were enhanced with Ti substitution up to 20 $\%$ compared to that of pure BiFeO$_3$ and Ti undoped Bi$_{0.9}$Gd$_{0.1}$FeO$_3$ samples. The enhanced magnetic properties might be attributed to the substitution induced suppression of spiral spin structure of BiFeO$_3$. An asymmetric shifts both in the field and magnetization axes of magnetization versus magnetic field (M-H) curves was observed. This indicates the presence of exchange bias effect in these compounds notably at room temperature.    \\

\end{abstract}

\maketitle
\section{Introduction} \label{I}
Multiferroic materials exhibit ferroelectric (or antiferroelectric)  properties in combination with ferromagnetic (or antiferromagnetic) properties in the same phase and have attracted considerable research interest due to their potential applications in multiple state memory elements, magnetic-field sensor and electric field controlled ferromagnetic resonance devices \cite{ref1,ref2,ref3,ref4,ref5,ref6,ref7}. Amongst the multiferroic ceramics being widely investigated, BiFeO$_3$ is a promising candidate for novel applications which allow mutual control of the electric polarization with a magnetic field and magnetization by an electric field \cite{ref6,ref7}. BiFeO$_3$  is ferroelectric below T$_C$ $\sim {~}$ 1103 K and antiferromagnetic below T$_N$ $\sim {~}$ 653 K, having rhombohedrally distorted perovskite ABO$_3$ (A = Bi, B= Fe) structure \cite{ref4}. Previous investigations revealed  that pure phase of BiFeO$_3$ is difficult to obtain \cite{ref8,ref9} and various impurity phases of  bulk BiFeO$_3$  have been reported, mainly comprising of Bi$_2$Fe$_4$O$_9$, Bi$_{36}$Fe$_{24}$O$_{57}$ and Bi$_{25}$FeO$_{40}$ \cite{ref10,ref11}.

In BiFeO$_3$, magnetic ordering is of antiferromagnetic type, having a spiral modulated spin structure (SMSS) with an incommensurate long-wavelength period of 62 nm \cite{ref12}. This spiral spin structure cancels the macroscopic magnetization and prevents the observation of the linear magnetoelectric effect \cite{ref13,ref14,ref15}. In addition, the bulk BiFeO$_3$  is characterized by serious current leakage problems due to the existence of a large number of charge centres caused by oxygen ion vacancies \cite{ref12}. These problems limit the use of BiFeO$_3$  for fabrication of multifunctional devices  \cite{ref16}. In order to overcome these problems, many attempts have been undertaken \cite{ref12}. Amongst which the partial substitution of Bi$^{3+}$ by ions such as Sm$^{3+}$ \cite{ref17,ref18}, Nd$^{3+}$ \cite{ref19,ref20} etc and also substitution of Fe$^{3+}$ by ions such as Cr$^{3+}$ \cite{ref21}, or simultaneous minor substitution of Bi$^{3+}$ and Fe$^{3+}$ by ions such as La$^{3+}$ and Mn$^{3+}$ or La$^{3+}$ and Ti$^{4+}$, respectively \cite{ref4, ref12} improved the magnetism and ferroelectricity in BiFeO$_3$. The partial substitution of Bi$^{3+}$ with ions having the biggest ionic radius has been found to effectively suppress the spiral spin structure of BiFeO$_3$ giving rise to the appearance of weak ferromagnetism \cite{ref22}. 

Among various kind of doping at A site, it was observed that the substitution of 10 $\%$ Gd in place of Bi enhanced the room temperature magnetization of BiFeO$_3$  \cite{ref6,ref27} as well as improved the phase purity of bulk BiFeO$_3$.
At B site, the partial substitution of Fe by Ti is found specially attractive as the substitution of Ti$^{4+}$ decreased the leakage current significantly and induced a remanent magnetization in BiFeO$_3$  \cite{ref11,ref26}. It is observed that at 200 K due to the substitution of Fe by Ti in BiFe$_{0.75}$Ti$_{0.25}$O$_3$, the area of the magnetization versus magnetic field (M-H) curve increased compared to that of pure BiFeO$_3$. However, at room temperature the M-H loop was still very narrow \cite{ref11} which indicates that substitution of 25 $\%$ Ti in place of Fe cannot improve notably the magnetic properties at room temperature (RT). Therefore, in the present investigation, we intend to study the effect of co-substitution of magnetic Gd and non-magnetic Ti in place of Bi and Fe, respectively, in BiFeO$_3$ samples. So far there are reports on the co-doping effect of La$^{3+}$ and Ti$^{4+}$ \cite{ref12}, La$^{3+}$ and Mn$^{3+}$ \cite{ref4}, La$^{3+}$ and V$^{5+}$ \cite{ref28}, respectively  in place of Bi and Fe in BiFeO$_3$. To the best of our knowledge there is no published data on dielectric and magnetic properties of Gd and Ti co-doped BiFeO$_3$ samples measured at RT. Therefore, nominal compositions of Bi$_{0.9}$Gd$_{0.1}$Fe$_{1-x}$Ti$_x$O$_3$ (x = 0.00-0.25) ceramics were synthesized by conventional solid state reaction technique and their structure, morphology, dielectric and magnetic properties were investigated. We observed that due to the combined effects of Gd and Ti, dielectric and magnetic properties of these ceramics were improved at RT. Moreover, an asymmetry exhibiting shifts both in the field and magnetization axes of M-H curves was also observed in Gd and Ti co-doped ceramics which indicates the presence of exchange bias (EB) effect in these compounds at RT.     

\section{Experimental details} \label{II}
The polycrystalline samples having compositions   Bi$_{0.9}$Gd$_{0.1}$Fe$_{1-x}$Ti$_x$O$_3$ (x = 0.00-0.25) were synthesized by using standard solid state reaction technique. The high purity oxides of Bi$_{2}$O$_{3}$, Gd$_{2}$O$_{3}$, Fe$_{2}$O$_{3}$, and TiO$_{2}$ powders were carefully weighed in stoichiometric proportion, mixed thoroughly with acetone and  grounded in an agate mortar until a homogeneous mixture was formed. The compacted mixtures of reagents taken in desired cation ratios were calcined at 800$^o$C for 1.5 hours in a programmable furnace. The calcined powders  were grounded again for 2 hours to get more homogeneous mixture. The powders were pressed into pellets of thickness 1 mm and diameter 12 mm by using a uniaxial hydraulic press and sintered at 825$^o$C for 5 hours at heating rate 10$^o$C per minute. 

The crystal structure of the samples (sintered powder) was determined from x-ray diffraction (XRD) data. XRD patterns were collected at RT using a diffractometer (Rigaku Ultimate VII) with CuK$_{\alpha}$  (${\lambda}$ = 1.5418 $\AA$) radiation. The microstructure of the surface of pellets was observed using a field emission scanning electron microscope (FESEM, JEOL, JSM 5800) equipped with the energy dispersive x-ray (EDX). The EDX analysis has been used to determine the overall chemical homogeneity and composition of the samples. The dielectric properties were measured at RT using an impedance analyzer (Agilent 4294 A) in the frequency range 100 Hz-20 MHz. The magnetic properties of the samples were characterized by using a Vibrating Sample Magnetometer (VSM, Lakeshore 7407 series) with a sensitivity of 10$^{-6}$ emu at RT.

\section{Results and discussions} \label{III}
\subsection{Structural characterizations} \label{I}
\begin{figure}[!hh]
\centering
\includegraphics[width=8.5cm]{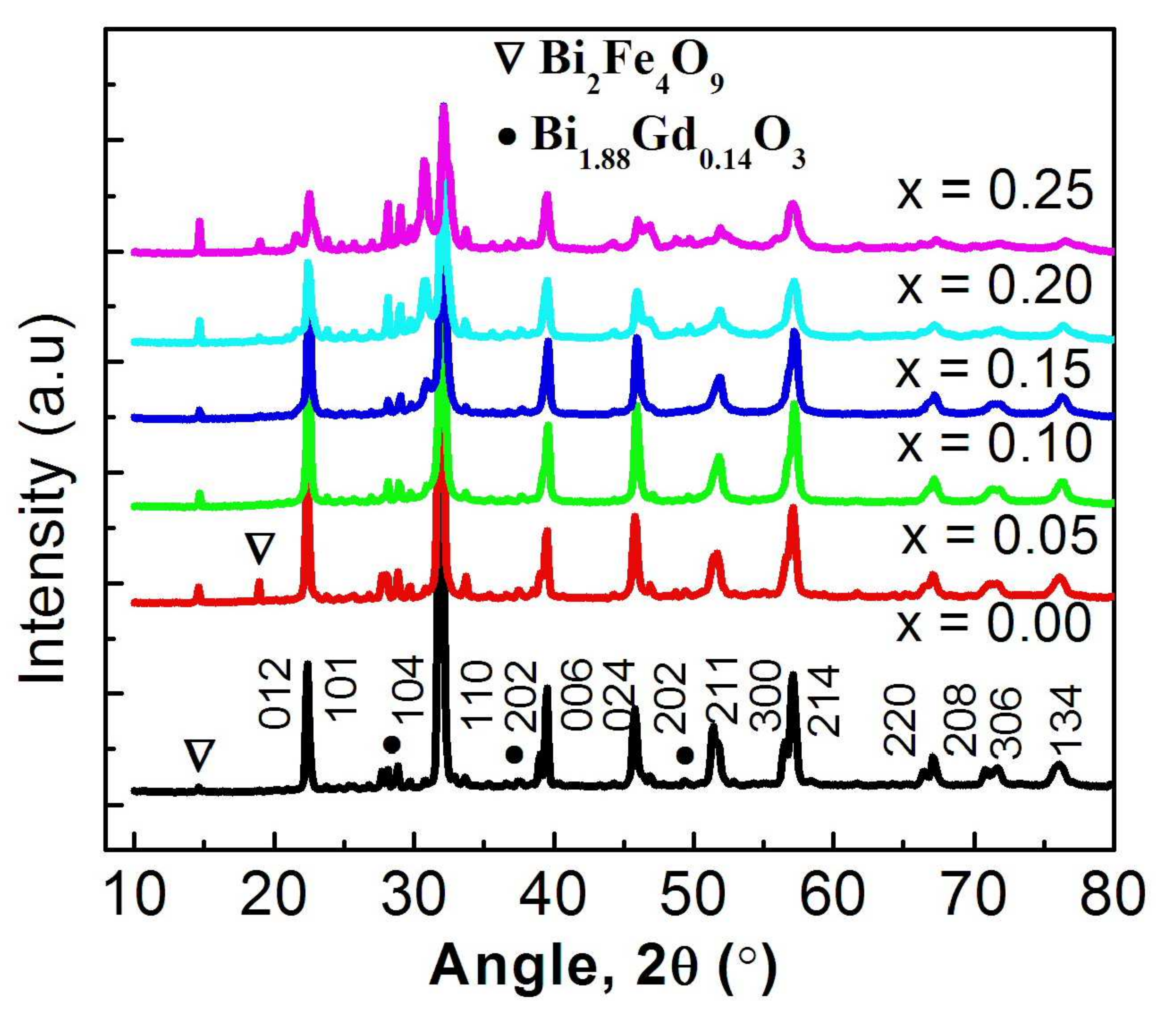}
\caption{X-ray diffraction patterns of Bi$_{0.9}$Gd$_{0.1}$Fe$_{1-x}$Ti$_x$O$_3$ (x = 0.00-0.25) ceramics.} \label{fig1}
\end{figure}
The XRD patterns of the Bi$_{0.9}$Gd$_{0.1}$Fe$_{1-x}$Ti$_x$O$_3$ (x = 0.00-0.25) ceramics sintered at temperature of their optimum density as shown in Fig. \ref{fig1} indicate the formation of polycrystalline structure. There are small traces of secondary phase Bi$_2$Fe$_4$O$_9$ as labeled by $\nabla$  still appear both in the undoped and doped samples. The apparently unavoidable formation of secondary phases during the solid-state synthesis of BiFeO$_3$ based materials have been reported in several articles \cite{ref12,ref72, ref11}. However, the presence of secondary phases do not affect ferroelectric and magnetic properties of the samples because these phases are neither magnetic nor ferroelectric at RT \cite{ref12}. The decrease in the peak intensities for x = 0.05 - 0.25 indicates the incorporation of Ti in Bi$_{0.9}$Gd$_{0.1}$Fe$_{1-x}$Ti$_x$O$_3$ ceramics \cite{ref12, ref74, ref75}. Furthermore, the twin peaks observed around 2${\theta}$ $\approx$ 40$^o$, 52$^o$ and 57$^o$ merges to form a broadened peak \cite{ref12, ref74} also confirms the substitution of Ti in these compounds. Splitting of peak around 2${\theta}$ $\approx$ 32$^o$ as Ti content increases to x =0.25 suggests a structure change from rhombohedral to orthorhombic \cite{ref76} and is consistent with that of reported articles \cite{ref77, ref78}.

\subsection{Morphological studies} \label{II}
\begin{figure}[!hh]
\centering
\includegraphics[width=5.75cm]{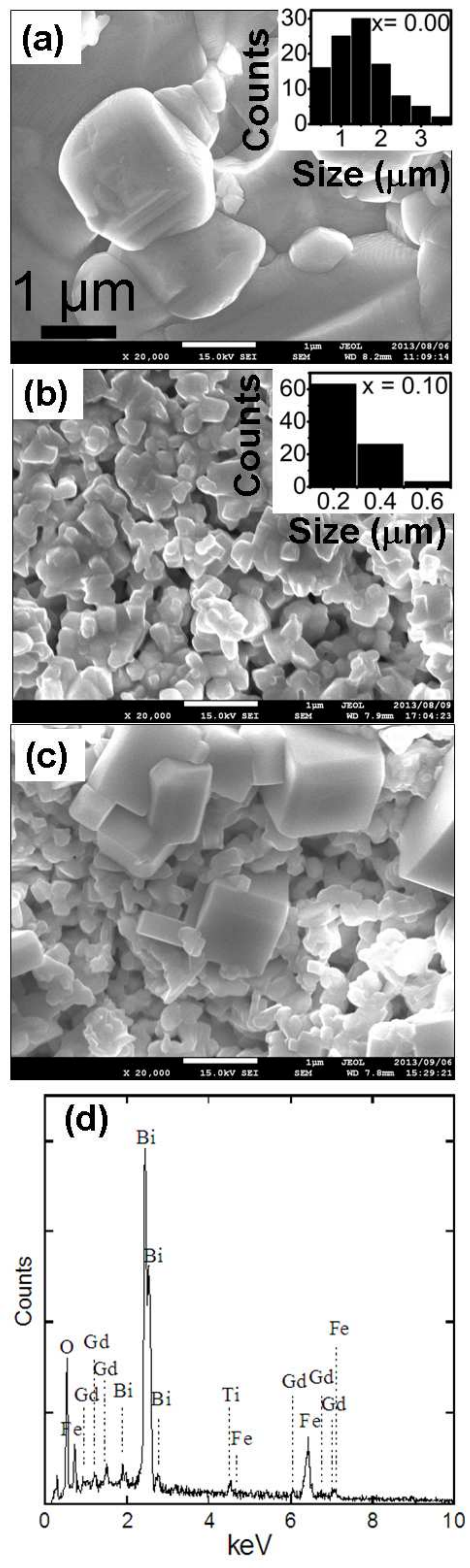}
\caption{FESEM micrograph's of Bi$_{0.9}$Gd$_{0.1}$Fe$_{1-x}$Ti$_x$O$_3$ ceramics: (a) x = 0.00, (b) x = 0.10 and (c) x = 0.25. Inset: respective histograms of images (a) and (b). Figure (d) shows an EDX pattern recorded from image (b).} \label{fig2}
\end{figure}
To investigate the microstructure of the surface of the pellets, FESEM imaging was carried out for all of the samples. Figure \ref{fig2}(a) demonstrates the surface morphologies of the pellets of the 10 $\%$ Gd doped Bi$_{0.9}$Gd$_{0.1}$FeO$_3$ sample (hereafter referred as Ti undoped Bi$_{0.9}$Gd$_{0.1}$FeO$_3$ sample). Notably, in the nominal compositions of Bi$_{0.9}$Gd$_{0.1}$Fe$_{1-x}$Ti$_x$O$_3$ ceramics, it is expected that titanium is in its tetravalent state and the non magnetic Ti$^{4+}$ substitutes Fe$^{3+}$. This is in fact not very simple because in order to accommodate a charge balance some Fe must be present as Fe$^{2+}$ or there must be present some type of non-stoichiometry. Figure \ref{fig2}(b) shows the microstructure of the Ti doped  Bi$_{0.9}$Gd$_{0.1}$Fe$_{0.9}$Ti$_{0.1}$O$_3$ ceramic with x = 0.10 as a representative of the x = 0.15 and 0.20 ceramics. Figure \ref{fig2}(c) is an FESEM image of 25 $\%$ Ti doped (the highest doping concentration) sample Bi$_{0.9}$Gd$_{0.1}$Fe$_{0.75}$Ti$_{0.25}$O$_3$ which also represents x = 0.05 ceramic. Insets of Figs. \ref{fig2}(a and b) show the histograms of the grain size distributions of the respective micrographs. It is clear from Figs. \ref{fig2}(a and b) and their respective histograms that the average grain size is reduced from $\sim {~}$1.5 $\mu m$ to $\sim {~}$200 $nm$ with increasing Ti concentration. Previous studies demonstrated that undoped BiFeO$_3$  had a dense microstructure with an average size of $\sim {~}$ 5 to $\sim {~}$ 15  $\mu m$  \cite{ref29,ref30}. In our investigation, we have also observed that the average grain size of pure BiFeO$_3$  is around $\sim {~}$ 5 $\mu m$ (data not shown here). Therefore, unlike the non-modified BiFeO$_3$ sample, the average grain size of the Ti undoped i.e. 10 $\%$ Gd doped Bi$_{0.9}$Gd$_{0.1}$FeO$_3$ sample is $\sim {~}$ 1.5 $\mu m$ which is well consistent with previous investigation \cite{ref27}. For the samples with x = 0.05, that is for an increase of 5 $\%$ Ti in Bi$_{0.9}$Gd$_{0.1}$Fe$_{1-x}$Ti$_x$O$_3$, a decreasing trend of the average grain size to $\sim {~}$ 200 $nm$ was observed, however, the grain distribution was not quite homogeneous rather there were couple of grains with larger sizes ($\sim {~}$ 1 $\mu m$) in this sample as was shown in Fig. \ref{fig2}(c). The further increase of the Ti content to 10 $\%$ (x = 0.10) reduced the average grain size to $\sim {~}$ 200 $nm$ as well as made the distribution of the grains homogeneous (Fig. \ref{fig2}(b)). Notably, increase of the Ti content from x = 0.10 to x = 0.15 or 0.20 does not reduce anymore the average grain size from $\sim {~}$ 200 $nm$. Due to the substitution of Ti, the average grain size was reduced and the peak in the XRD pattern of the Ti substituted sample was widened as shown in Fig. \ref{fig1}.

However, for 25  $\%$ Ti doping, the average grain size remain unchanged, that is $~$ 200 $nm$ but some of the grains grow to larger sizes (($\sim {~}$1 $\mu m$) (Fig. \ref{fig2}(c)). Previous investigation suggested that the grain growth depends upon the concentration of oxygen vacancies \cite{ref31} and diffusion rate of the ions. Large number of oxygen vacancies are generate in pure BiFeO$_3$ due to highly volatile nature of Bi. An increment of the Ti doping concentration in Bi$_{0.9}$Gd$_{0.1}$Fe$_{1-x}$Ti$_x$O$_3$ decreased the average grain size significantly due to the fact that Ti possesses a higher valence than Fe and suppresses the formation of oxygen vacancies. Therefore, the significant reduction of the average grain size in Gd and Ti co-doped BiFeO$_3$ samples could be interpreted by the suppression of the formation of oxygen vacancies because of the requirements of the charge compensation. The decreased oxygen vacancies lead to a lower grain growth rate which is actually a consequence of slower oxygen ion motion \cite{ref32}. To ascertain the level of doping in these samples, EDX analysis was done on all of them. Figure \ref{fig2}(d) is an EDX spectrum of the Bi$_{0.9}$Gd$_{0.1}$Fe$_{0.9}$Ti$_{0.1}$O$_3$ sample taken from image \ref{fig2}(b) as a representative of all other samples. This EDX spectrum of the Bi$_{0.9}$Gd$_{0.1}$Fe$_{0.9}$Ti$_{0.1}$O$_3$ ceramic shown in Fig. \ref{fig2}(d) confirms the presence of expected amounts of Bi, Gd, Fe, Ti and O in the synthesized sample.

\subsection{Dielectric measurements} \label{III}
\begin{figure}[!hh]
\centering
\includegraphics[width=8cm]{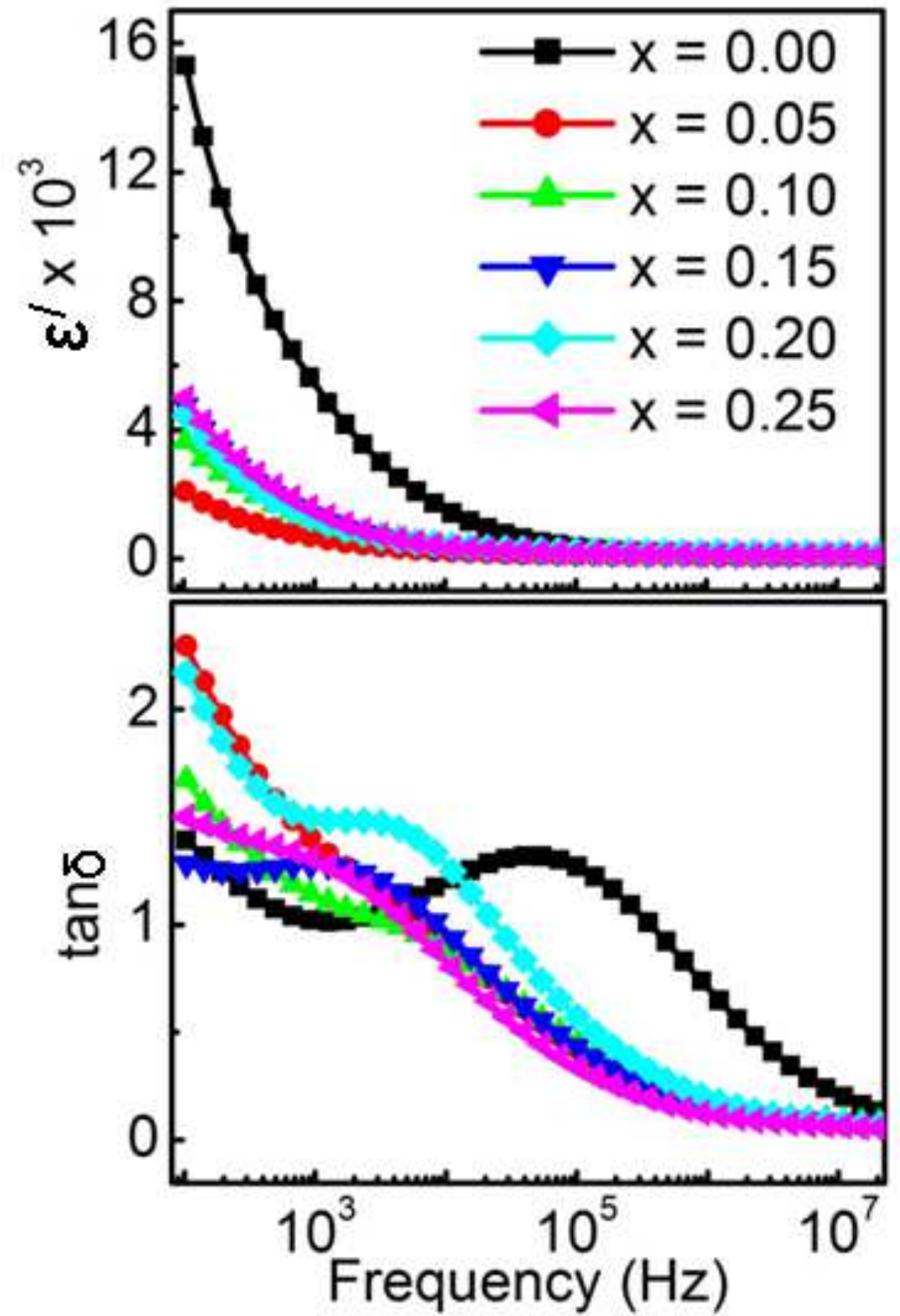}
\caption{(a) Semi-log plot of dielectric constant of Bi$_{0.9}$Gd$_{0.1}$Fe$_{1-x}$Ti$_x$O$_3$ ceramics : (a) x = 0.00, (b) x = 0.05 (c) x = 0.10 (d) x = 0.15, (e) x = 0.20 and (f) x = 0.25 recorded at RT in the wide frequency range from 100 Hz - 20 MHz. (b) Variation in tan $\delta$ as a function of frequency at RT.} \label{fig6}
\end{figure}

Figure \ref{fig6}(a) illustrates the frequency dependence of the dielectric constant ($\in^/$) of Bi$_{0.9}$Gd$_{0.1}$Fe$_{1-x}$Ti$_x$O$_3$ ceramics with (a) x = 0.00 (b) x = 0.05 (c) x = 0.10 (d) x = 0.15 (e) x = 0.20 and (f) x = 0.25 measured at RT in the frequency range 100 Hz - 20 MHz. The variation of dielectric loss tangent (tan $\delta$) in the same frequency range is shown in Fig. \ref{fig6}(b).  

\begin{table}[!h]
\caption{Dielectric constant at different frequencies for x = 0.00-0.25. The table shows that in a wide range of high frequencies (50 kHz to 20,000 kHz) the dielectric constant amongst co-doped ceramics is higher for x =0.10, 0.15 and 0.20 compared to that of x =0.05 and 0.25 samples.}  
\begin{center}
\begin{tabular}{|l|l|l|l|l|l|l|}
 \hline
{Frequency (kHz)} &\multicolumn{6}{c|}{Dielectric constant for x =} \\
\cline{2-7}
 &0.00&0.05&0.10&0.15&0.20&0.25\\
   
\hline
50&552&139&244&244&285&211\\
\hline
100&367&120&209&210&253&178\\
\hline
1000&142&91&154&158&205&128\\
\hline
10000&99&82&136&139&187&113\\
\hline
20000&96&81&134&137&184&112\\
\hline
\hline

\end{tabular}
\end{center}
\end{table}

The dielectric constant of Ti undoped sample Bi$_{0.9}$Gd$_{0.1}$FeO$_3$ (x = 0.00) is maximum at the lower frequencies which decreases sharply with increasing frequency up to about 300 kHz and then becomes almost constant at higher frequencies $>$300 kHz. Gd and Ti co-doped Bi$_{0.9}$Gd$_{0.1}$Fe$_{1-x}$Ti$_x$O$_3$ (x = 0.05-0.25) ceramics also demonstrate slight dispersion at low frequency and a frequency independent behavior in  a wide range of high frequencies. This low frequency dispersion is common in dielectric and ferroelectric materials \cite{ref61} and explained in the light of space charge polarization as discussed by Maxwell \cite{ref62} and Wagner \cite{ref63}. At low frequencies, the space charges are able to follow the frequency of the applied field, while at a high frequency they may not have time to build up and undergo relaxation. The observed large values of dielectric constant at low frequencies for Bi$_{0.9}$Gd$_{0.1}$Fe$_{1-x}$Ti$_x$O$_3$ ceramics (x = 0.00-0.25) attributed to the interfacial polarization due to high concentration of oxygen vacancies. The oxygen vacancies are produced due to highly volatile nature of Bi, and /or multiple oxidation states of Fe \cite{ref64, ref65} as was also mentioned earlier. 
The value of the frequency dependent dielectric constant may be described in two-step. Firstly, the dielectric constant is decreased due to Ti substitution compared to that of Ti undoped sample. For the smallest amount of Ti dopant (05 $\%$) introduced in Bi$_{0.9}$Gd$_{0.1}$Fe$_{1-x}$Ti$_x$O$_3$, the dielectric constant at 100 kHz drops to almost one eighth of its original value for Ti undoped Bi$_{0.9}$Gd$_{0.1}$FeO$_3$ sample. The dielectric constant at low frequencies are much smaller for all Gd and Ti co-doped samples Bi$_{0.9}$Gd$_{0.1}$Fe$_{1-x}$Ti$_x$O$_3$ (x =0.05-0.25) compared to that of Ti undoped sample Bi$_{0.9}$Gd$_{0.1}$FeO$_3$. 

Secondly, an increment of Ti doping concentration from 05 $\%$ to 10, 15 and 20 $\%$ increases the dielectric constant and then decreases again for 25 $\%$ doping concentration. The dielectric constants of Bi$_{0.9}$Gd$_{0.1}$Fe$_{1-x}$Ti$_x$O$_3$ (x = 0.00-0.25) ceramics at high frequencies (50 kHz - 20 MHz) are inserted in table I. From the table it is clear that in a wide range of high frequencies (50 kHz to 20 MHz), the dielectric constant of Gd and Ti co-doped samples with x =0.10, 0.15 and 0.20 is higher than that of co-doped samples with x =0.05 and 0.25. It is also noteworthy that the dielectric constant of Gd and Ti co-doped ceramics with x =0.10, 0.15 and 0.20 is higher at any frequency than that of pure BiFeO$_3$ ceramics for which dielectric constant is around 100 as reported in Ref. \cite{ref66}. The higher values of the dielectric constants at high frequencies for samples with x =0.10, 0.15 and 0.20 could be attributed to the reduced grain size of these compounds. The reduced grain size ultimately increased volume fraction of the grain boundaries leading to high dielectric constant. The dielectric constant was comparatively lower for samples with x =0.05 and 0.25 although the average grain size was also smaller than that of Ti undoped sample. This might be due to the fact that the grain size distribution was not homogeneous for these two samples and there was grain growth of some of the grains to $\sim {~}$1 $\mu m$ (Fig. \ref{fig2}(c)).

The  frequency dependent region of dielectric constant decreased due to Ti substitution compared to that of Ti undoped sample. Due to 05 $\%$ Ti doping, the frequency dependent region of the dielectric constant of sample Bi$_{0.9}$Gd$_{0.1}$Fe$_{0.95}$Ti$_{0.05}$O$_3$ is reduced to 30 kHz from 300 kHz of Ti undoped sample Bi$_{0.9}$Gd$_{0.1}$FeO$_3$. The frequency dependent region of all other Gd and Ti co-doped samples are also decreased compared to that of Ti undoped sample. This indicates that the extrinsic behavior (low frequency behavior) of these co-doped ceramics is suppressed with Ti doping concentration. It was already mentioned that Ti possesses a higher valence than Fe and the substitution of Ti  suppresses the formation of oxygen vacancies because of the requirements of the charge compensation. The phenomenon of oxygen vacancies has direct correspondence with the leakage current density \cite{ref29}. From the  measurements of leakage current densities (data not shown here), we have observed that for a substitution of 5 $\%$ Ti in place of Fe, leakage current density is reduced significantly than that of Ti undoped sample. For a further increment of Ti concentration, leakage current densities in the Gd and Ti co-doped samples were increased, however, their values were always lower than that of Ti undoped sample Bi$_{0.9}$Gd$_{0.1}$FeO$_3$. It is expected that leakage current densities in the Gd and Ti co-doped samples were decreased due to the low concentration of charge defects such as oxygen vacancies \cite{ref29,ref65}, and therefore, the low frequency dispersion was suppressed due to Ti substitution. 

The dielectric loss tangent also shows frequency dependency as illustrated in Fig.\ref{fig6}(b). At low frequency, the dielectric loss is lower for Ti undoped sample Bi$_{0.9}$Gd$_{0.1}$FeO$_3$ compared to that of Gd and Ti co-doped samples. Interestingly, at high frequency, the loss tangent is higher for Ti undoped sample Bi$_{0.9}$Gd$_{0.1}$FeO$_3$ and lower for all Gd and Ti co-doped samples. Due to the low ($<$ 0.15) loss tangent values at higher frequencies, the Gd and Ti co-doped samples might have potential applications in high-frequency microwave devices.

\subsection{Magnetic characterization} \label{IV}
\begin{figure}[!hh]
\centering
\includegraphics[width=8.5cm]{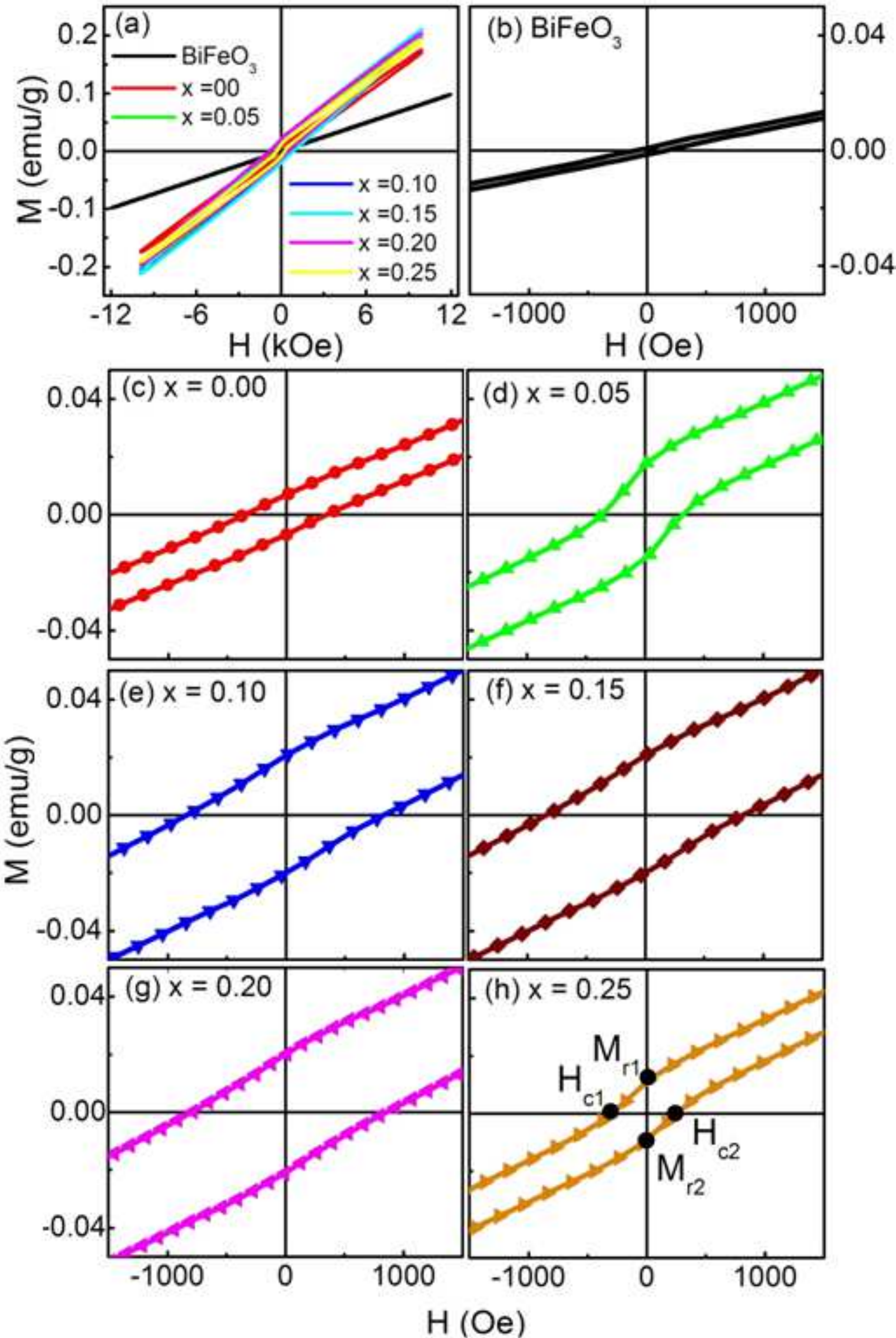}
\caption{(a) M-H hysteresis loops of pure BiFeO$_{3}$  and  Bi$_{0.9}$Gd$_{0.1}$Fe$_{1-x}$Ti$_x$O$_3$ (x =0.00-0.25) ceramics at room temperature. (b) An enlarged view of the low-field M-H hysteresis loop of pure BiFeO$_{3}$. (c-h) An enlarged view of the low-field M-H hysteresis loops of Bi$_{0.9}$Gd$_{0.1}$Fe$_{1-x}$Ti$_x$O$_3$ samples obtained at RT : (c) x = 0.00 (d) x = 0.05 (e) x = 0.10 (f) x = 0.15 (g) x = 0.20 and (h) x = 0.25. Figure (h) also shows the left and right side coercive fields (Hc$_1$ and Hc$_2$) and remanent magnetizations (Mr$_1$ and Mr$_2$) of M-H loops. An asymmetric shifts both in the field and magnetization axes of Gd and Ti co-doped samples (d-h) indicates the existence of the EB phenomenon.} \label{fig4}
\end{figure}

Magnetization experiments were carried out at RT using VSM. Figure \ref{fig4} (a) shows M-H curves for pure BiFeO$_3$, Ti undoped Bi$_{0.9}$Gd$_{0.1}$FeO$_3$, and Gd and Ti co-doped Bi$_{0.9}$Gd$_{0.1}$Fe$_{1-x}$Ti$_x$O$_3$ (x = 0.05-0.25) ceramics measured at RT with an applied magnetic field of up to $\pm$12 kOe. All the samples show unsaturated magnetization loops which confirm the basic antiferromagnetic nature of the compounds. As is seen from Fig. \ref{fig4} (b) that pure BiFeO$_3$  sample possess a very narrow hysteresis loop with a very small but non-zero remanent magnetization (0.0009 emu/g) and a coercive field of $\sim {~}$110 Oe at RT. This is due to the fact that pure BiFeO$_3$ is antiferromagnetic (AFM) which does not possess any spontaneous magnetization \cite{ref66} but has a residual magnetic moment for a canted spin structure. 


\begin{figure}[!hh]
\centering
\includegraphics[width=7cm]{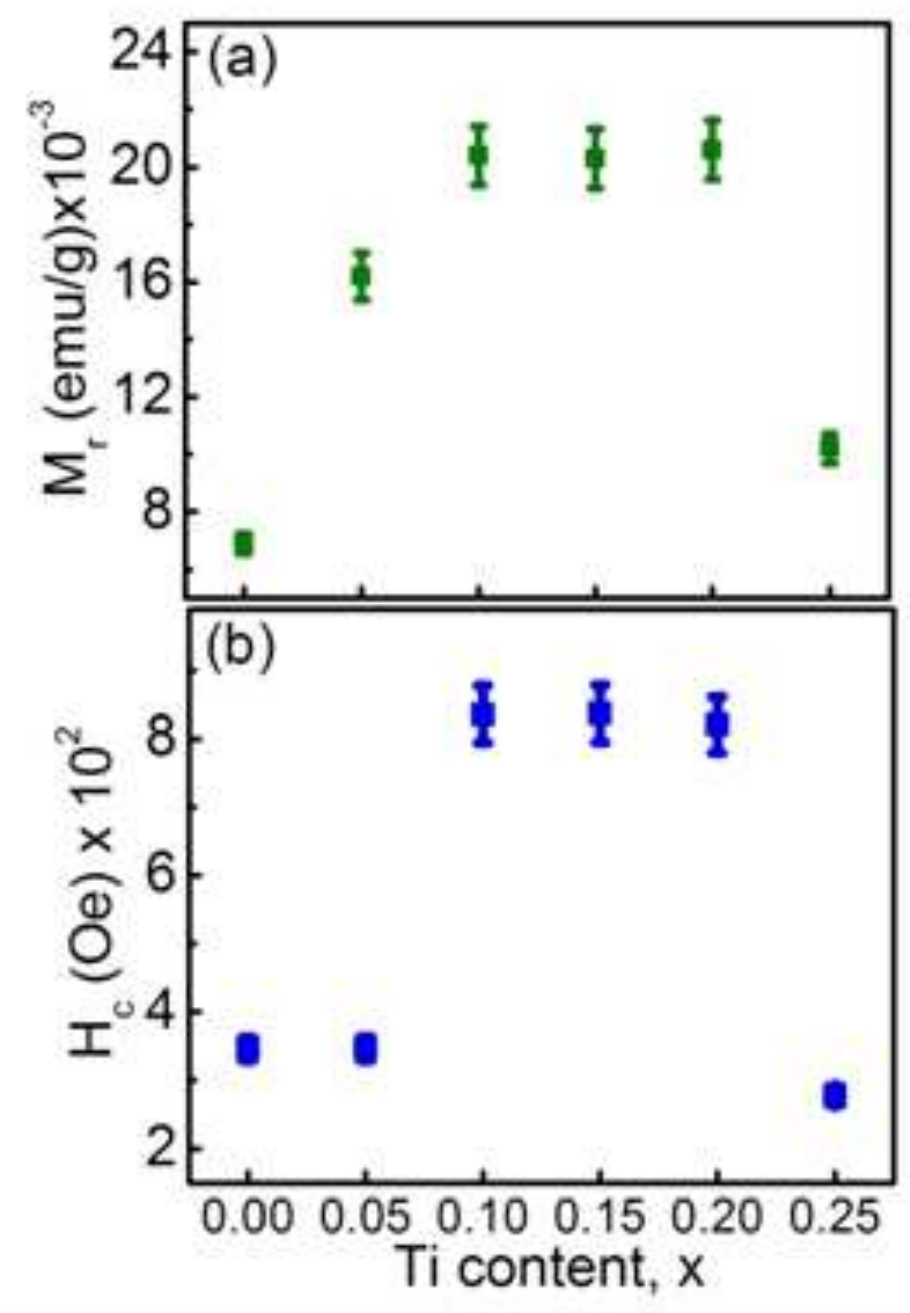}
\caption{(a,b) Variation of remanent magnetization and coercive fields at RT, respectively, in Bi$_{0.9}$Gd$_{0.1}$Fe$_{1-x}$Ti$_x$O$_3$ (x = 0.00-0.25) as a function Ti concentration.} \label{fig5}
\end{figure}

Figures \ref{fig4} (c-h) demonstrate an enlarged view of the low-field M-H hysteresis loops of Bi$_{0.9}$Gd$_{0.1}$Fe$_{1-x}$Ti$_x$O$_3$ ceramics measured at RT with (c) x = 0.00 (d) x = 0.05 (e) x = 0.10 (f) x = 0.15, (g) x = 0.20 and (h) x= 0.25. From Figs. \ref{fig4} (c-h) a non-zero remanent magnetization and coercive field are observed. The remanent magnetization (M$_r$) is defined as M$_{r}$ = $|$(M$_{r1}$-M$_{r2}$)$|$/2 where M$_{r1}$ and M$_{r2}$ are the magnetization with positive and negative points of intersection with H = 0, respectively \cite{ref4} as shown in Fig. \ref{fig4} (h). The coercive field (H$_c$) is given by H$_c$ = $|$(H$_{c1}$-H$_{c2}$)$|$/2, where H$_{c1}$ and H$_{c2}$ are the left and right coercive fields \cite{ref33, ref34}, respectively as shown also in Fig. \ref{fig4} (h). Calculated values of M$_{r}$ and H$_c$ were plotted in Figs. \ref{fig5} (a and b) respectively for Bi$_{0.9}$Gd$_{0.1}$Fe$_{1-x}$Ti$_x$O$_3$ (x =0.00-0.025) ceramics. An increment of the 05 $\%$ Ti doping concentration in place of Fe in Bi$_{0.9}$Gd$_{0.1}$Fe$_{1-x}$Ti$_x$O$_3$ increased the remanent magnetization at RT, although the coercivity remained unchanged as shown in Figs. \ref{fig5} (a) and (b). A further increase in Ti doping concentration to 10 $\%$ dramatically enhanced M$_r$ and H$_c$ (Figs. \ref{fig5} (a) and (b)). For any further increment of the Ti doping concentration from 10 $\%$ to 15 and 20 $\%$ the remanent magnetizations and coercive fields were unaltered (Figs. \ref{fig5} (a) and (b)) and highest. Due to 10 $\%$ Gd doping in BiFeO$_3$ improved magnetic property was observed (Fig. \ref{fig4} (c)) as was reported earlier in Ref. \cite{ref27}. This enhancement of the  magnetization at RT was attributed \cite{ref27} to the structural distortion in the perovskite with change in Fe$-$O$-$Fe angle. It was expected that this structural distortion could lead to suppression of the spin spiral and hence enhanced the magnetization in the Bi$_{0.9}$Gd$_{0.1}$FeO$_3$ system. 

As in the case of 10 $\%$ Gd doped Bi$_{0.9}$Gd$_{0.1}$FeO$_3$, the hysteresis loops of Gd and Ti co-doped ceramics were not really saturated at 12 kOe, however, a significant enhancement in remanent magnetization and coercive field was observed with the increase in Ti doping concentration in the Bi$_{0.9}$Gd$_{0.1}$Fe$_{1-x}$Ti$_x$O$_3$ system. In a similar investigation, room temperature magnetic properties were carried out for only Ti doped BiFe$_{1-x}$Ti$_{x}$O$_3$ ceramics with  varying x up to 0.35 \cite{ref76}. In that investigation, the highest remanent magnetization was 0.0095 emu/g \cite{ref76} for 20 $\%$ Ti doped (not co-doped) sample BiFe$_{0.80}$Ti$_{0.20}$O$_3$. In the case of presently investigated Gd and Ti co-doped ceramics Bi$_{0.9}$Gd$_{0.1}$Fe$_{1-x}$Ti$_x$O$_3$ with x = 0.10. 0.15, 0.20, the remanent magnetization is 0.02 emu/g, which is double than that of only Ti doped BiFe$_{0.80}$Ti$_{0.20}$O$_3$ ceramic. However, there is a possibility that the observed magnetic behavior could be due to the presence of magnetic secondary phases which would not be detected by XRD technique because of its limit of detection. 

The coercive field is also higher for Gd and Ti co-doped samples Bi$_{0.9}$Gd$_{0.1}$Fe$_{1-x}$Ti$_x$O$_3$ with x = 0.10. 0.15, 0.20 compared to that of only Ti doped BiFe$_{1-x}$Ti$_{x}$O$_3$ (x= 0.00-0.35) ceramics observed in Ref. \cite{ref76}. Hence, it is clear that Gd and Ti co-doped Bi$_{0.9}$Gd$_{0.1}$Fe$_{1-x}$Ti$_x$O$_3$ ceramics exhibit enhanced magnetic properties at RT showing weak ferromagnetic (WFM) antiferromagnetism. The origin of WFM \cite{ref35} due to a change in bond angle of Fe$-$O$-$Fe as a result of distortion created by Ti doping. The statistical distribution of Fe$^{3+}$ and Fe$^{2+}$ (created due to charge compensation) ions in the octahedral also lead to an increase in magnetization and weak ferromagnetism. However, for 25 $\%$ Ti doping in Bi$_{0.9}$Gd$_{0.1}$Fe$_{1-x}$Ti$_x$O$_3$, the remanent magnetization and coercive field both again decreased and the antiferromagnetic order typical to BiFeO$_3$ parent phase is likely to be recovered. This might be associated with the structure phase transition from rhombohedral to orthorhombic at x = 0.25 \cite{ref76}.

Notably, from FESEM images (Fig. \ref{fig2} (b)) the average grain size was smallest ($\sim {~}$200 $nm$) for Gd and Ti co-doped samples with x =0.10, x =0.15 and x =0.20 for which the remanent magnetization and coercive fields are the highest as shown in Figs. \ref{fig5} (a) and (b). The dielectric constant over a wide range of high frequencies (50 kHz to 20 MHz) was also highest for these three samples compared to that of the other two samples. Therefore, the observed results clearly indicate that the dielectric and magnetic properties of Gd and Ti co-doped samples are related to the microstructure of the compositions.
  
Beside this enhanced room temperature magnetic properties in  Gd and Ti co-doped Bi$_{0.9}$Gd$_{0.1}$Fe$_{1-x}$Ti$_x$O$_3$ ceramics, the most intriguing feature observed from the enlarged view (Figs. \ref{fig4}(d-f)) of the low-field M-H hysteresis loops is their asymmetric behavior. An asymmetry exhibiting shifts both in the field and magnetization axes can be clearly observed for all the samples. This indicates the presence of exchange bias (EB) effect in these compounds. The exchange bias effect usually occurs in ferromagnetic and antiferromagnetic bilayers or multilayers in which the two coercive fields of the magnetic hysteresis loop are not symmetric, and the centre of the magnetic hysteresis loop shifts to the left or right \cite{ref33, ref34, ref36}. Recent investigations also demonstrate that the exchange bias effect can also exist in compounds or composites which allow the coexistence of both a ferromagnetic component and an antiferromagnetic component \cite{ref33, ref36, ref37}. In compounds like NdMnO$_3$ \cite{ref36} or La$_{1-x}$Pr$_x$CrO$_3$ \cite{ref37}, the exchange bias effect is different from what appears in bilayer and other interface structures.

\begin{table}[!h]
\caption{The table shows the calculated values of H$_{EB}$ and M$_{EB}$ for Bi$_{0.9}$Gd$_{0.1}$Fe$_{1-x}$Ti$_x$O$_3$ ceramics observed at RT. Table also demonstrates comparison of the H$_{EB}$  for La$_{0.8}$Bi$_{0.2}$Fe$_{1-x}$Mn$_x$O$_3$ multiferroics observed at 200 K \cite{ref4} and  Bi$_{0.9}$Gd$_{0.1}$Fe$_{1-x}$Ti$_x$O$_3$ ceramics observed at RT.}  
\begin{center}
\begin{tabular}{|l|l|l|l|l|}
 \hline
{Ti} &\multicolumn{2}{c|}{Bi$_{0.9}$Gd$_{0.1}$Fe$_{1-x}$Ti$_x$O$_3$}& \multicolumn{2}{c|}{La$_{0.8}$Bi$_{0.2}$Fe$_{1-x}$Mn$_x$O$_3$} \\
concentration &\multicolumn{2}{c|}{ceramics at}& \multicolumn{2}{c|}{multiferroics at} \\
(x) &\multicolumn{2}{c|}{RT }& \multicolumn{2}{c|}{ 200 K \cite{ref4}} \\
\cline{2-5}
 &H$_{EB}$&M$_{EB}$ x 10$^{-4}$&Mn&H$_{EB}$ \\
&(Oe) &emu/g&concentration (x) &(Oe)\\

\hline
0.00&$00\pm 0.00$&$0.00\pm 00$&0.2&$+24$\\
\hline
0.05&$16\pm 0.8$&$15\pm 0.8$&$0.3$&$+08$\\
\hline
0.10&$06\pm 0.3$&$04\pm 0.2$&$0.4$&$-37$\\
\hline
0.15&$06\pm 0.3$&$05\pm 0.25$&$--$&$--$\\
\hline
0.20&$-36\pm 02$&$-05\pm 0.25$&$--$&$--$\\
\hline
0.25&$26\pm 1.3$&$09\pm 0.45$&$--$&$--$\\
\hline
\end{tabular}
\end{center}
\end{table}

The EB field H$_{EB}$ is generally defined
as H$_{EB}$ = -(H$_{c1}$+H$_{c2}$)/2 where H$_{c1}$ and H$_{c2}$ are the left
and right coercive fields, respectively as shown by solid circles in Fig. \ref{fig4} (h) \cite{ref33, ref34}. The asymmetry in the magnetization axes is defined as M$_{EB}$ = (M$_{r1}$+M$_{r2}$)/2 where M$_{r1}$ and M$_{r2}$ are the magnetization with positive and negative points of intersection with H = 0, respectively \cite{ref4}. Calculated values of H$_{EB}$ and M$_{EB}$ for different doping concentrations observed at RT are inserted in table II. The exchange bias phenomena was also reported in a similar multiferroics La$_{0.8}$Bi$_{0.2}$Fe$_{1-x}$Mn$_x$O$_3$ as a function of Mn doping at 20 K, 100 K and 200 K \cite{ref4}. The presence of EB phenomenon in these compounds was assumed to be associated with induced exchange anisotropy at the interface between FM and AFM phases \cite{ref4}. For a comparison the calculated values of H$_{EB}$ observed at 200 K for multiferroics La$_{0.8}$Bi$_{0.2}$Fe$_{1-x}$Mn$_x$O$_3$ \cite{ref4} have been displayed in table II along with the values observed at RT for Bi$_{0.9}$Gd$_{0.1}$Fe$_{1-x}$Ti$_x$O$_3$ (x =0.00-0.25) ceramics. From the table it is evident that the exchange bias fields vary randomly as a function of Ti doping in Bi$_{0.9}$Gd$_{0.1}$Fe$_{1-x}$Ti$_x$O$_3$ ceramics as was also observed  in La$_{0.8}$Bi$_{0.2}$Fe$_{1-x}$Mn$_x$O$_3$ \cite{ref4} due to Mn doping. It is expected that the EB in these system originate from the exchange interaction at the interfaces of weakly ferromagnetic and antiferromagnetic components \cite{ref38}. The asymmetry in the M-H curve was also observed at RT with 10 $\%$ Sm doping in a similar type of ceramics Bi$_{1-x}$Sm$_x$FeO$_3$ \cite{ref39}. The authors reported that in this compound while most of the sample remain largely antiferromagnetic, regions in the Sm-doped BiFeO$_3$ sample become weakly ferromagnetic and their interaction results an asymmetric behaviour in the M-H curve although the effect was not observed for 20 $\%$ Sm doping. However, in the present investigation, EB bias effect was observed at RT up to 25 $\%$ Ti doping.\\

\section{Conclusions} \label{II}
The nominal compositions of Bi$_{0.9}$Gd$_{0.1}$Fe$_{1-x}$Ti$_x$O$_3$ ( x =0.00 -0.25) ceramics were synthesized and their structural, morphological, dielectric and magnetic properties at room temperature were investigated. The combined effects of Gd and Ti doping in BiFeO$_3$ inhibit grain growth and stabilize dielectric constant over a wide range of high frequencies by suppressing dispersion at low frequencies. At high frequencies, dielectric loss attained minimum values which indicates the potential applicability of Gd and Ti co-doped BiFeO$_3$ ceramics in high-frequency microwave devices. Due to the substitution of Ti in  Bi$_{0.9}$Gd$_{0.1}$Fe$_{1-x}$Ti$_x$O$_3$ (x =0.05 -0.25) ceramics, the remanent magnetizations and coercive fields were enhanced at room temperature.  This enhancement of the room temperature magnetization might be attributed to the structural distortion in the perovskite with change in Fe$-$O$-$Fe angle due to Ti doping. Our investigation revealed that for Gd and Ti co-doped  Bi$_{0.9}$Gd$_{0.1}$Fe$_{1-x}$Ti$_x$O$_3$ ceramics with x =0.10, 0.15 and 0.25, the average grain size is smallest, the dielectric constant at high frequencies is highest, the remanent magnetizations and coercive fields are also highest compared to that of the compounds with x =0.05 and 0.25. Therefore, it may be concluded that in these co-doped ceramics  10 to 20 $\%$ Ti doping is optimum to have better multiferroic properties. We are carrying out further investigations in detail on these optimized ceramics. The presence of exchange bias effect in these compounds at room temperature is another fascinating feature which was also reported in similar systems \cite{ref4,ref33, ref36, ref37} although at low temperature.

\section{Acknowledgements}
We sincerely acknowledge University Grants Commission (UGC), Dhaka, Bangladesh for providing financial support.

\end{document}